\documentclass[twocolumn]{aastex63}
\usepackage{graphics,amsmath,amsbsy}
\usepackage{lineno}

\newcommand{\lapprox} {\, \lower3pt\hbox{$\sim$}\llap{\raise2pt\hbox{$<$}}\,}
\newcommand{\gapprox} {\, \lower3pt\hbox{$\sim$}\llap{\raise2pt\hbox{$>$}}\,}

\definecolor{mrkred}{RGB}{160,0,0}

\definecolor{mrk}{RGB}{0,0,160}

\renewcommand{\deg}{\protect{^{\circ}}}

\shorttitle{Sizes of metric solar radio sources}
\shortauthors{Gordovskyy et al.}
\begin{document}
\title{Sizes and shapes of sources in solar metric radio bursts}

\author[0000-0003-2291-4922]{Mykola Gordovskyy}
\affiliation{Department of Physics \& Astronomy, University of Manchester, Manchester M13 9PL, UK}

\author[0000-0002-8078-0902]{Eduard P. Kontar}
\affil{School of Physics \& Astronomy, University of Glasgow, Glasgow, G12 8QQ, UK}

\author[0000-0003-1967-5078]{Daniel L. Clarkson}
\affil{School of Physics \& Astronomy, University of Glasgow, Glasgow, G12 8QQ, UK}

\author[0000-0002-4389-5540]{Nicolina Chrysaphi}
\affil{LESIA, Observatoire de Paris, Université PSL, CNRS, Sorbonne Université, Université de Paris, 5 place Jules Janssen, 92195 Meudon, France}
\affil{School of Physics \& Astronomy, University of Glasgow, Glasgow, G12 8QQ, UK}

\author[0000-0002-7089-5562]{Philippa K. Browning}
\affil{Department of Physics \& Astronomy, University of Manchester, Manchester M13 9PL, UK} 

\begin{abstract}
Metric and decametric radio-emissions from the Sun are the only direct source of information about the  dynamics of non-thermal electrons in the upper corona. In addition, the combination of spectral and imaging (sizes, shapes, and positions) observations of  low-frequency radio sources can be used as a unique diagnostic tool to probe plasma turbulence in the solar corona and inner heliosphere. The geometry of the low-frequency sources and its variation with frequency are still not understood, primarily due to the relatively low spatial resolution available for solar observations. Here we report the first detailed multi-frequency analysis of the sizes of solar radio sources observed by the Low-Frequency Array (LOFAR). Furthermore, we investigate the source shapes by approximating the derived intensity distributions using 2D Gaussian profiles with elliptical half-maximum contours. These measurements have been made possible by a novel empirical method for evaluating the instrumental and ionospheric effects on radio maps based on known source observations. The obtained deconvolved sizes of the sources are found to be smaller than previous estimations, and often show higher ellipticity. The sizes and ellipticities of the sources inferred using 2D Gaussian approximation, and their variation with frequency are consistent with models of anisotropic radio-wave scattering in the solar corona.
\end{abstract}

\section{Introduction}
\label{s-intro}

The plasma density and magnetic field in the upper solar corona and inner heliosphere are not sufficient to produce detectable bremsstrahlung hard X-ray or gyrosynchrotron emissions. Hence, metric and decametric coherent radio-emissions are the sole source of information about energetic electrons in these layers of the solar atmosphere \citep{mcla85,dule98,pivi08,klee05,reie11}. This information is vital for understanding the underlying mechanisms of electron acceleration and transport in the corona, and escape of energetic electrons from the corona into the heliosphere.  

Metric and decametric radio emissions from the Sun are characterised by a myriad of various bursts with complicated frequency and spatial structure varying on tens of millisecond time-scales \cite[e.g.][]{wadu69, kund82, bare94, shae18, kuko19, chre20, mage20}. Observing them requires sub-second temporal resolution in many narrow frequency bands. At the same time, spatial resolution is also extremely important, because knowing the locations and sizes of the low-frequency sources is essential for characterising the physical properties of the emitting region, as well as the conditions in the corona and inner heliosphere, where radio emission propagates. 

Radio-waves in this spectral range are strongly affected by scattering and refraction in the turbulent solar corona \cite[e.g.][]{bost77, bast94, kone17}. Recent years saw significant progress in low-frequency radio observation of the Sun and their theoretical interpretation. It has been shown that sizes, locations, and temporal evolution of solar radio burst sources can be used as an important diagnostic tool for plasma turbulence in the solar corona and inner heliosphere \citep{kone17,chre18,gore19,kone19,chee20}. This has re-ignited interest to the question of the source sizes in solar radio bursts.  However, although solar radio bursts have been intensively studied for the past six decades, there is no certainty regarding the sizes of the emission sources \cite[e.g.][]{1985srph.book..289S}. This is because the relatively small interferometric baselines usually available for deca-metric solar radio observations, in combination with the need for high temporal cadence, result in a relatively low spatial resolution, making the evaluation of geometric properties of the sources very challenging.

There was a number of studies of the solar emission source sizes in the frequency range of 100 MHz--1 GHz. For instance, \citet{mere06,mere15} used the combinations of Nancay Radioheliograph (NRH) and Giant Meterwave Radio Telescope (GMRT) to map the radio-emission from non-flaring Sun and found that between frequencies of 200--400~MHz the emission sources corresponding to active regions can be as little as 0.5~arcmin. However, only a handful of single-frequency observational estimations of source sizes in solar radio bursts are available at lower frequencies below 100MHz. \citet{abre76,abre78} found that the sizes of sources corresponding to individual stria in type III bursts normally have sizes between 20--40~arcmin at 25~MHz. \citet{chsh78} found similar sizes between about 25--40~arcmin in type III, as well as type II, IV and V bursts observed at 26.4~MHz. More recently, \citet{kone17} used the Low-Frequency Array \cite[LOFAR, ][]{vane13} tied-array beam (TAB) data to investigate  type III sources observed close to the centre of solar disk. It was shown that at 32~MHz the major axis of the half-maximum contour of the source is about 19~arcmin. \citet{mure21} used LOFAR data obtained in the interferometric mode to study a type III source observed off the solar limb. They found that at 35~MHz its shape can be approximated by an ellipse with the major and minor axes of about 18 and 10~arcmin, respectively.

LOFAR is a state-of-the-art instrument which offers unprecedented opportunities to map solar radio emission with very high spectral and temporal resolution. In the TAB mode it provides data with high frequency resolution (12~kHz) and temporal cadence (10~ms). However, the 3.5~km baseline normally used for solar TAB observations results in a relatively large point-spread function (PSF) or `dirty beam', which is further affected by factors other than the instrument's baseline, such as the Earth's ionosphere. Hence, it is important to analyse LOFAR imaging data using an empirically-obtained PSF, based on known source observations.

\begin{table}[h!]
  \begin{center}
    \caption{Characteristics of the considered solar events, showing date and time, location in the sky (azimuth and zenith distance), and a brief description of the event. Azimuth $A$ and elevation $z$ are in degrees.}
    \begin{tabular}{l l l l l} 
\hline
      Date & Time, UT & $A$ & $z$ & Type of event\\
\hline
      16/04/2015 & 11:55 & 188 & 47 & Single type III\\
      27/04/2015 & 12:25 & 200 & 49 & Single type III\\
      06/05/2015 & 11:48 & 188 & 53 & Double type III\\
      20/06/2015 & 12:01 & 198 & 60 & Type IV\\
      25/06/2015 & 11:46 & 185 & 60 & Type II/IV\\
      12/07/2017 & 08:52 & 118 & 46 & Type III storm\\
      13/07/2017 & 07:42 & 101 & 36 & Single type III\\
      15/07/2017 & 11:03 & 164 & 58 & Type II, III\\
      09/09/2017 & 11:17 & 176 & 42 & Type IV?\\
      \hline
    \end{tabular}
  \end{center}
\end{table}

\begin{figure}[ht!]
\centering{\includegraphics[width=0.48\textwidth]{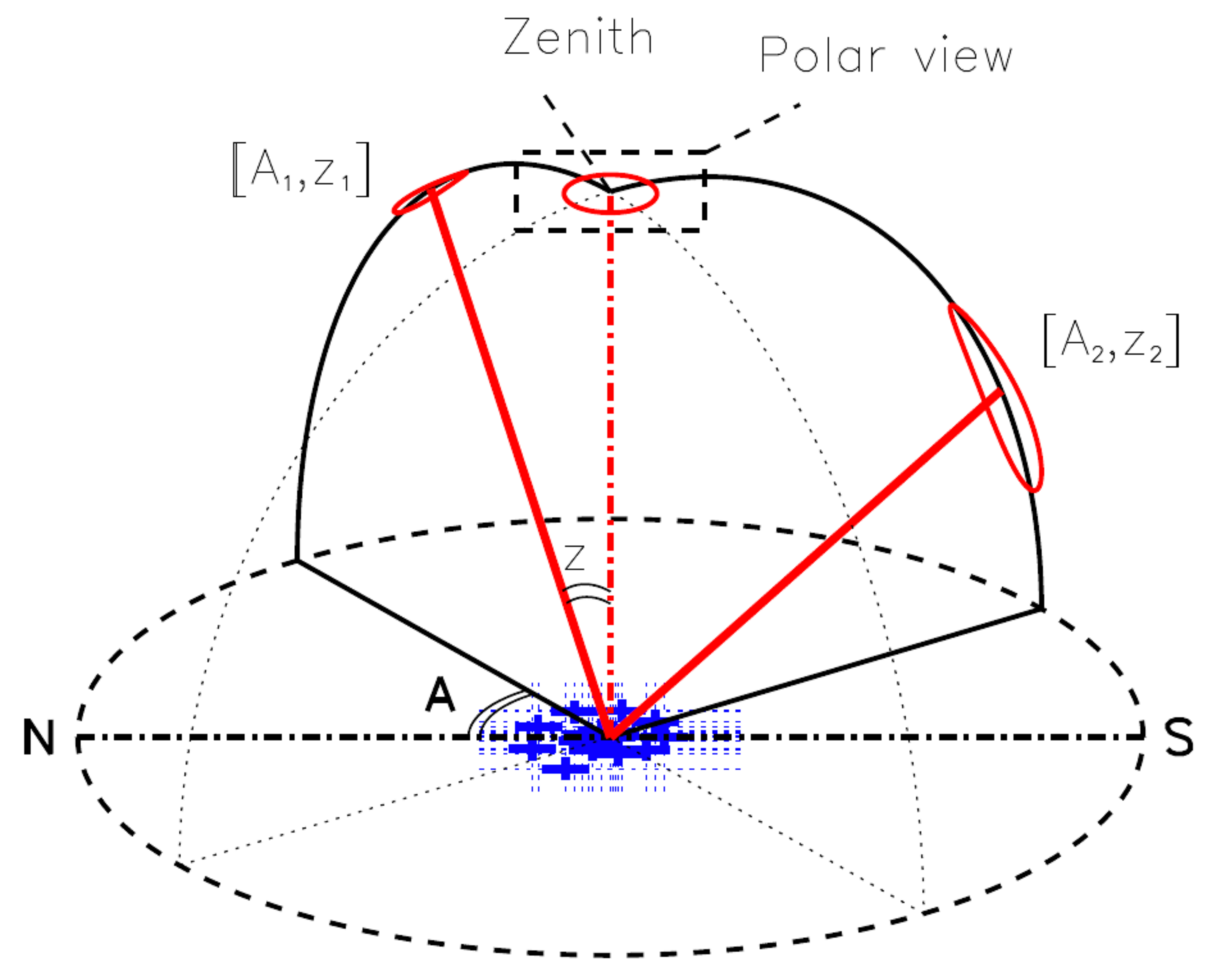}}
\caption{Sketch demonstrating the PSF translation procedure and the main parameters used in it.}
\label{f-sketch}
\end{figure}

In this study, we develop a novel method to evaluate the effective PSF of a radio array using observations of a known source (i.e. a radio calibrator) at an arbitrary location. The method is then used to correct the intensity maps of solar radio bursts observed by LOFAR in the frequency range 30-45~MHz, and evaluate the sizes and shapes of sources observed in nine randomly chosen events. 

\section{Data and its analysis}
\label{s-data}

\subsection{Observations}
\label{s-obs}

In this study we investigate nine randomly chosen solar radio bursts in the frequency range 30-45~MHz observed in 2015 and 2017. Spectral imaging data for these events have been obtained using LOFAR in the tied-array beam mode. The considered solar radio bursts are of different types (Table~1), ranging from short-duration events containing a single type III burst (e.g. 13/7/2017 event) to long-duration type II and IV storms (e.g. 20/6/2015 and 12/7/2017 events). At the same time, characteristics of their dynamic spectra, such as the relatively narrow-band emission ($\Delta f/f \approx 0.1-0.3$) and fast frequency drift of individual elements of dynamic spectra, indicate that the emission in all considered events is produced by the plasma mechanism \citep{gizh58}. Some of these bursts have been studied in more detail in recent years \cite[e.g.][]{kone17,chre18,chre20,chee20}). In each of the studied solar events we chose a moment corresponding to a bright feature in the dynamic spectrum.

In addition to solar radio data, in order to evaluate the effective PSF of LOFAR, the ionospheric refraction at different frequencies, and estimate the value of ionospheric shift, we consider observations of a calibrator, i.e. a known compact radio-source. Namely, we use 24 individual Tau A spectral imaging observations obtained in the same frequency range, 30--45~MHz, using the same LOFAR observing mode. These 24 observations were obtained in July--September 2017 with the object at different coordinates $[A,z]$, with elevations over the horizon (or altitudes) ranging from 15$\deg$ to 60$\deg$. 

For solar spectral-imaging data obtained in April-June 2015, 169 beams have been used, while for solar and Tau A observations during July-September 2017, 217 beams have been used. The average spacing between beams was about 400~arcsec. 

The data obtained for the nine solar radio bursts and for the 24 individual Tau A observations has been pre-processed using the same procedure. Firstly, the data has been degraded to temporal resolution of 1.6~s and to a frequency resolution of 320~kHz. From that point the data is treated individually for 16 frequency channels, from 30 to 45~MHz with 1~MHz step. The integration time for imaging is 2~s, both for solar and Tau~A observations.

For each frequency channel, a constant background $I_b$ has been subtracted from the data, with $I_b$ calculated as the average between the faintest $N$ beams, with $N$ being equal to half the number of available beams. For instance, for the data defined using 217 beams, the 109 faintest beams are selected and the average value between them is assumed to be $I_b$. This is based on the assumption that in at least half of the field of view area, the real signal is not greater than the noise.

\begin{figure*}[ht!]
\centering{\includegraphics[width=0.9\textwidth]{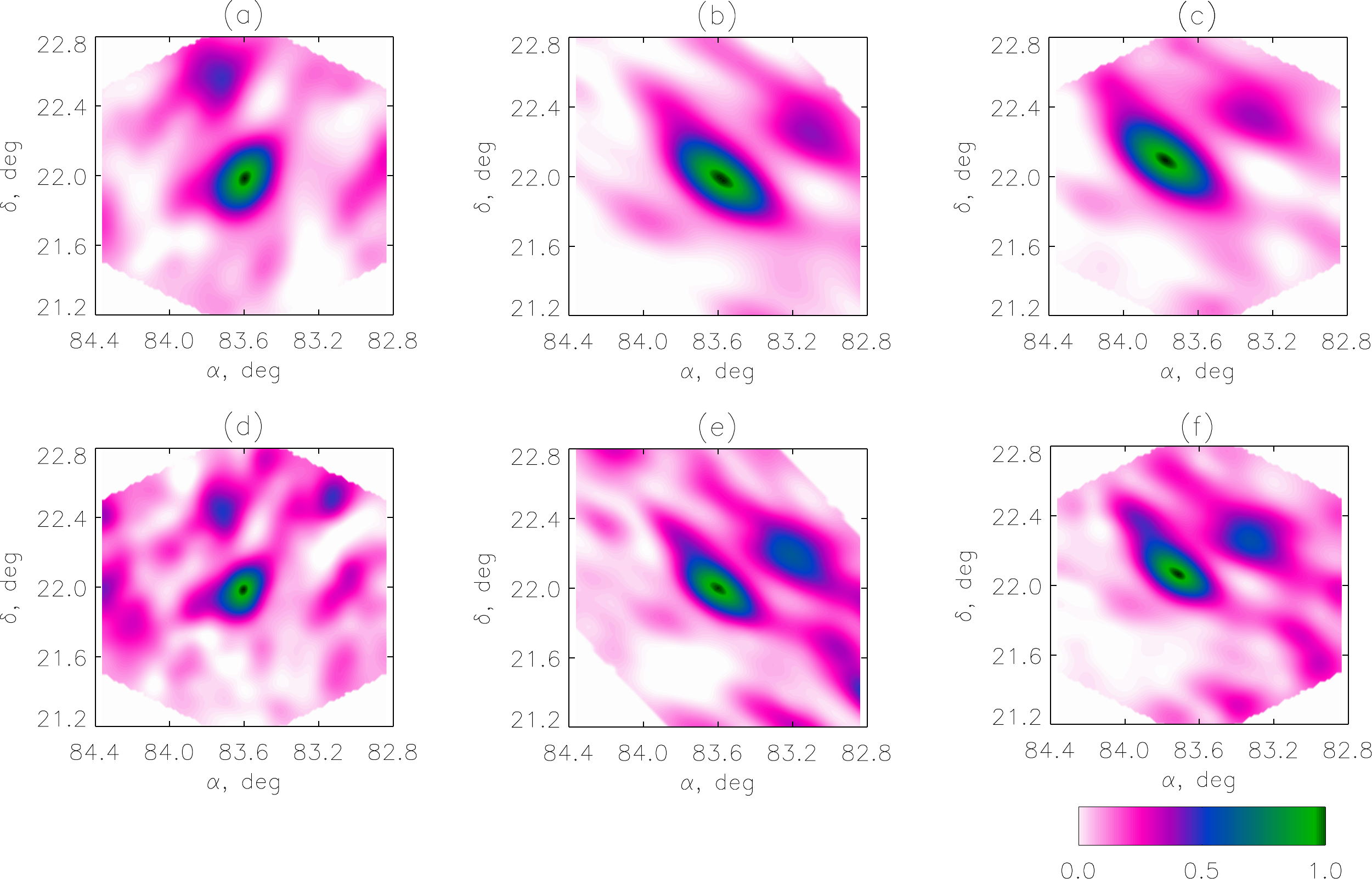}}
\caption{Left panels (a and d) show Tau A observed by LOFAR at 07:00~UT on 13 July 2017. Middle panels (b and e) show these intensity maps translated to the location where Tau A is observed at 10:05~UT on 09/09/2017. Right panels (c and f) show actual intensity maps of Tau A observed at 10:05~UT on 09/09/2017. Upper panels (a-c) correspond to the frequency 30~MHz, while lower panels correspond to 45~MHz.}
\label{f-taua}
\end{figure*}

\begin{figure*}[ht!]
\centering{\includegraphics[width=0.9\textwidth]{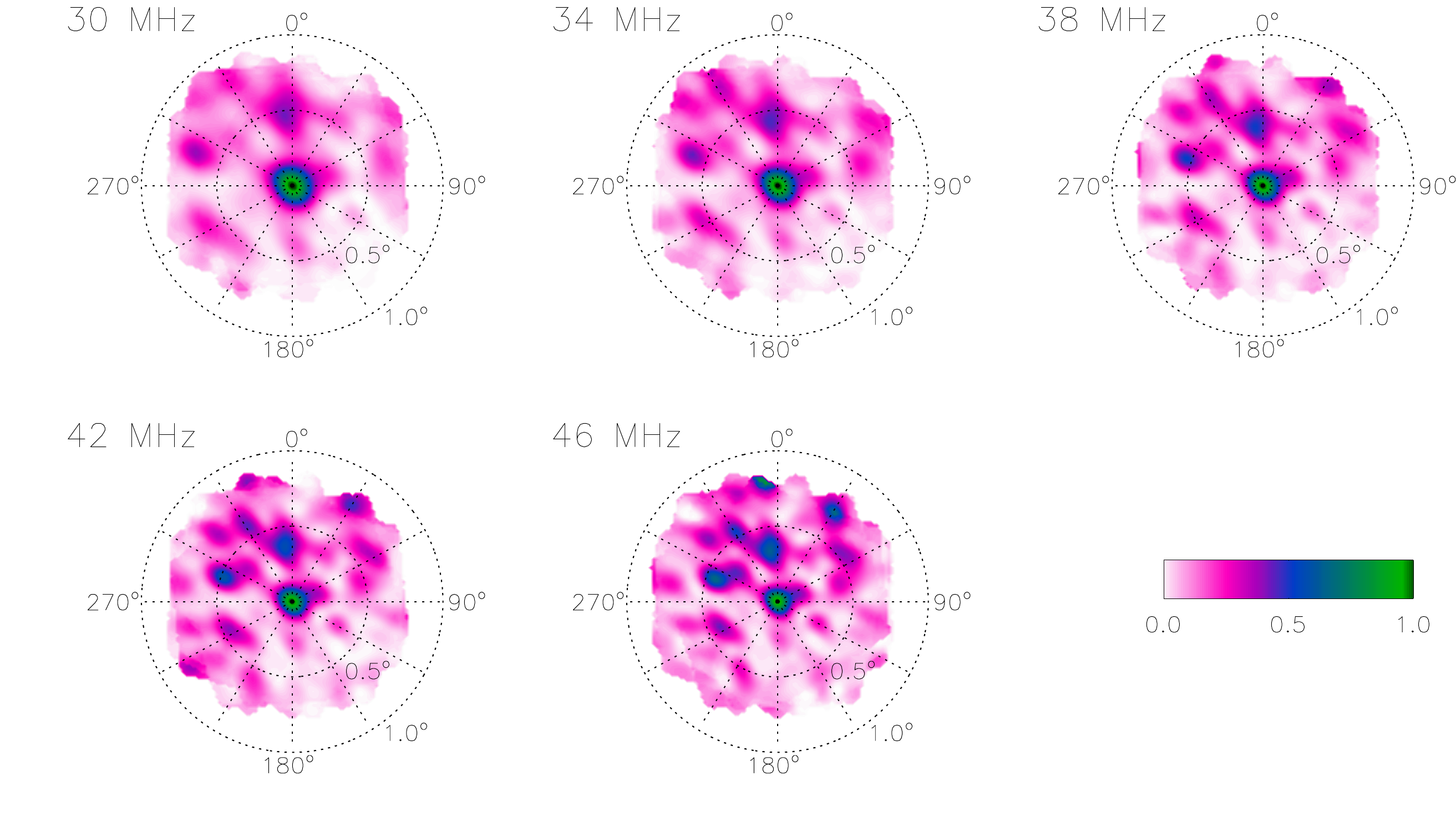}}
\caption{LOFAR PSF based on 22 Tau A observations projected to zenith above the LOFAR core. Frequencies are shown above the panels.}
\label{f-psf}
\end{figure*}

The initial analysis of spectral-imaging data described above yields apparent radio intensity maps $I_d(f,A,z)$, which are affected by instrumental and ionospheric effects. In the first approximation, the apparent maps $I_d(f,A,z)$ can be considered as convolutions of the real maps $I_0 (f,A,z)$ with the point-spread function (PSF) $\mathcal{F}(f,A,z,\Delta A,\Delta z)$, and shifted by $\Delta z_\mathrm{ir}(f)$ due to radio-wave refraction in the Earth's ionosphere:

\begin{eqnarray}
I_d(f,&A&,z) = \nonumber \\
&I_0&(f,A,z-\Delta z_\mathrm{ir}(f)) \star \mathcal{F}(f,A,z,\Delta A,\Delta z),
\label{eq-main}
\end{eqnarray}
where $A$ and $z$ are the azimuth and zenith distance ($z=90\deg - a$, where $a$ is the elevation). Therefore, to evaluate the actual location and shape of the observed object $I_0 (f,A,z)$, one needs to deconvolve the dirty map, reducing it for the effect of the PSF, and shift it by $\Delta z_\mathrm{ir}$. For this purpose, we empirically evaluate the effective PSF of LOFAR (Section~\ref{s-psf}) and the value of ionospheric shift at these frequencies (Section~\ref{s-ionos}).

In this study it is assumed that the source intensity distributions are represented by 2D Gaussian profiles with elliptical half-maximum contours:
\begin{eqnarray}
U(x,y)=U_0\exp&&\left(-\frac{\left[(x-x_0)\cos\theta+(y-y_0)\sin\theta\right]^2}{a^2} - \right. \nonumber \\
&&\left. \frac{\left[(y-y_0)\cos\theta-(x-x_0)\sin\theta\right]^2}{b^2}\right), \nonumber
\end{eqnarray}
where $U_0$ is the amplitude, $x_0$ and $y_0$ are coordinates of the maximum, $2\sqrt{\ln2}\,a$ and $2\sqrt{\ln2}\,b$ are the major and minor axes of the ellipse corresponding to the half-maximum contour, and $\theta$ is the tilt angle of that ellipse.
Obviously, the real intensity distributions are likely to be more complex than 2D Gaussian distributions. We make this assumption in order to regularise the observed and derived intensity distributions and avoid spurious results.

\subsection{Effective PSF of LOFAR in tied-array beam mode}
\label{s-psf}

The observed maps of Tau A are convolutions of its actual shape and the LOFAR PSF (or `dirty beam') (Equation~\ref{eq-main}). Since the actual shape of Tau A is known, one can evaluate the shape of the PSF at different frequencies by deconvolving dirty Tau A maps, yielding the effective PSF shapes for the locations $[A,z]$, where Tau A was observed. 

Deconvolution of Tau A maps has been performed using an iterative algorithm based on the modified method of Burger and van Cittert described  by \citet{jane66}. For the apparent intensity distribution $I_d$ formed by the convolution of the actual intensity distribution $I_0$ and the PSF $F$
\[
I_d = I_0 \star F
\]
with $I_0$ known, the first iteration for the PSF $F$ can be found as 
\[
F_1 = 2 I_d - I_d \star I_0.
\]
Each following iteration is calculated as
\[
F_{P+1} = F_P + I_d - F_P \star I_0.
\]
The iterations are performed until the difference between $P$th and $(P+1)$th iterations is negligible. Since the apparent shape of Tau A observed by LOFAR is much bigger than the actual Tau A shape, and therefore its detailed structure is not important, for the purpose of this deconvolution the actual Tau A shape ($I_0$) is represented by a 2D Gaussian. Parameters of the analytical function representing the Tau A intensity distribution are obtained by fitting a 2D Gaussian shape to the Tau A intensity map from \citet{mago79}.

Obviously, in order to deconvolve the images of solar events, the PSF needs to be determined for the location of the Sun at the time of the event. To do this, we have developed a PSF translation procedure, which makes it possible to evaluate the PSF for an arbitrary location in the sky based on the shape of PSF in another arbitrary location. This procedure is based on the assumption that all antennas of the interferometer are located in the horizontal plane and the locations are fixed.

Assume that the PSF corresponding to a source located at zenith has a form

\[
\mathcal{P}_z = f(A,z), 
\]
where $A$ and $z$ are the azimuth and zenith distance, respectively (Figure~\ref{f-sketch}). This PSF function can be rewritten as a function of small angles $\Delta A$ and $\Delta z$ as 
\begin{equation}
\mathcal{P}_z= f(A_1,\Delta A,\Delta z),
\label{eq-psf1}
\end{equation}
where $\Delta A= z \sin(A-A_1)$ and $\Delta z = z \cos(A-A_1)$, and $A_1$ is some arbitrary azimuth. As the size of the object (i.e. values of $\Delta A$ and $\Delta z$)  is assumed to be small, the small angle approximation can be used. 

For an object observed away from zenith, the PSF will expand in the zenith angle direction because of the effective baseline contraction due to the projection effect:
\begin{equation}
\mathcal{P}_1= f(A_1,z_1,\Delta A,\Delta z/\cos(z)). 
\label{eq-psf2}
\end{equation}
For small values of $\Delta A$ and $\Delta z$ and $z_1 \gg \Delta z$, the full azimuth and zenith angle can be calculated for every point on the maps as 
\begin{eqnarray}
A &=& A_1+ \Delta A / \sin(z) \\
\label{eq-psf3}
z &=& z_1 + \Delta z.
\label{eq-psf4}
\end{eqnarray}

Therefore, if the PSF of an instrument is known at zenith, Equations~\ref{eq-psf1}--\ref{eq-psf4} can be used to predict the PSF for any arbitrary location in the sky $A_1,z_1$, and vice versa. This, in turn, means that if the PSF is known for a location $A_1,z_1$, it can be evaluated for another arbitrary location $A_2,z_2$. Since the apparent intensity distribution corresponding to a point-like source should be equal to the PSF of an instrument, observing a point-like source in an arbitrary location in the sky makes it possible to evaluate the PSF for any other arbitrary location in the sky using Equations~\ref{eq-psf1}--\ref{eq-psf4}.

\begin{figure*}[ht!]
\centering{\includegraphics[width=0.6\textwidth]{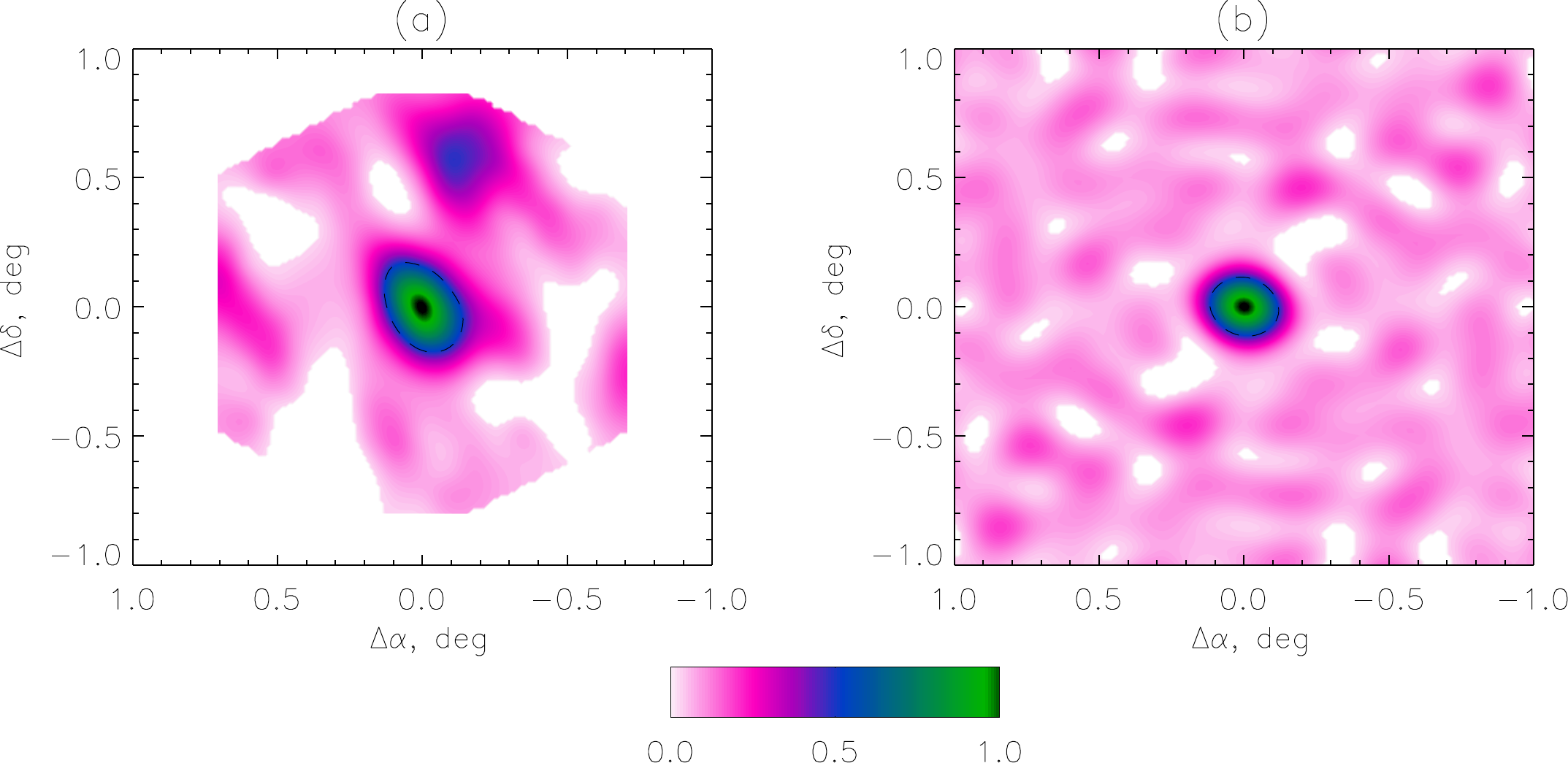}}
\caption{Tau A observed by LOFAR at 30MHz at 07:00~UT on 13 July 2017 compared with the nominal PSF of LOFAR calculated for the same frequency, time and sky coordinates.}
\label{f-nominal}
\end{figure*}

\begin{figure}[ht!]
\centering{\includegraphics[width=0.38\textwidth]{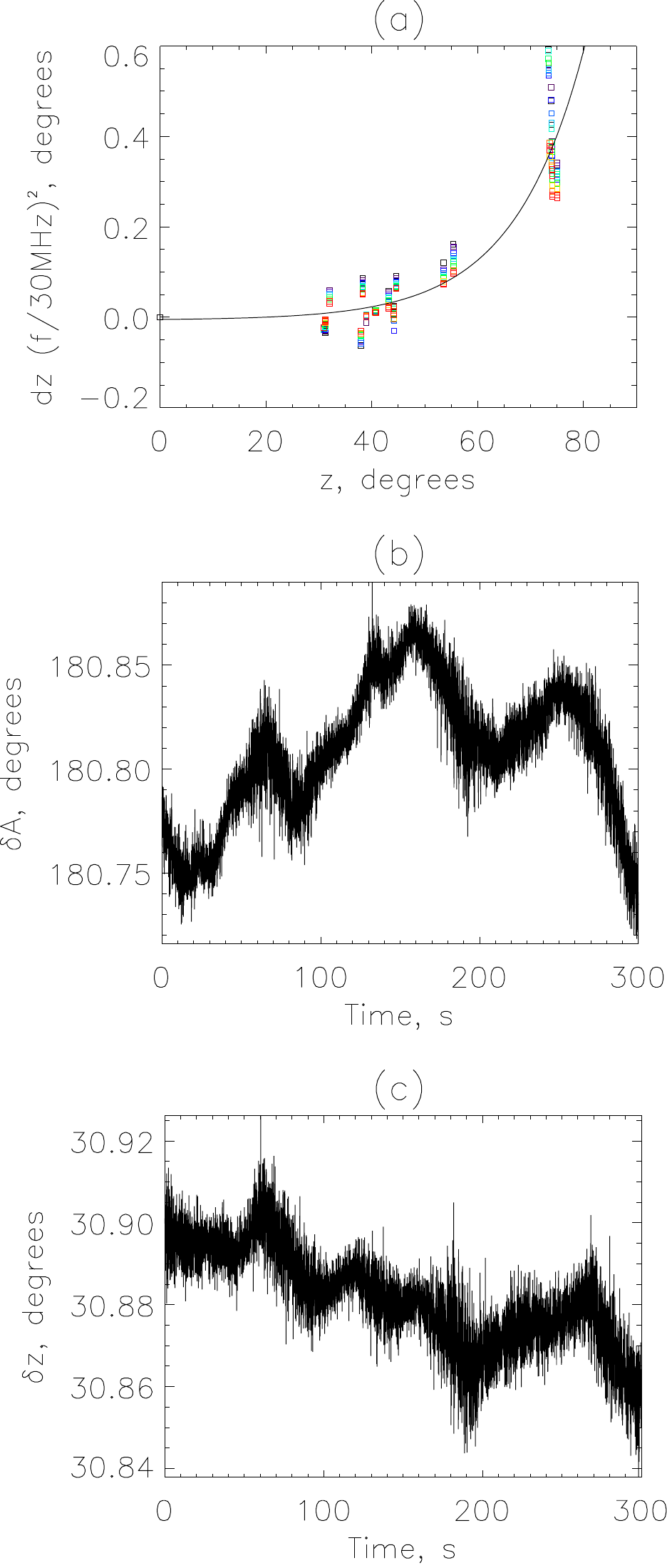}}
\caption{Panel (a): Zenith distance off-sets of Tau A centroids observed at different zenith distances. The values are normalised by $f^2$ factor to remove the effect of frequency variation (see Section~\ref{s-ionos}). Different colours represent different frequencies from 30 MHz (black) to 48 MHz (red). The black line shows the analytical approximation for the $\mathcal{R}(z)$ function. Panels (b) and (c): Apparent azimuth and zenith distance of Tau A  observed at 30~MHz with the motion due to the Earth rotation removed. Horizontal axis shows the time after 11:01~UT on 12 July 2017.} 
\label{f-ionos}
\end{figure}

In order to test the developed method, we compare the PSFs derived using Tau A observations in one location in the sky with maps of Tau A observed in a different location. Figure~\ref{f-taua} shows Tau A maps observed on the 9th of September 2017 about 36$\deg$ over the horizon and PSFs derived for this particular location in the sky based on Tau A observed about 45$\deg$ over the horizon on the 13th July 2017 (the intrinsic size of Tau A is significantly smaller than the observed sources and therefore, the maps shown in Figure~\ref{f-taua} are determined primarily by the instrumental PSF). Although the Tau A maps observed on 13 July and 9 September are very different, Tau A maps observed on 9 September and the PSFs calculated for this particular observation based on 13 July observations are very similar, both in terms of the mainlobe shapes and locations, and relative intensities of the sidelobes. 

The above analysis demonstrates that the PSF translation procedure works well for LOFAR observations in TAB mode. It also demonstrates that the quality of calculated PSFs heavily depends on the quality of the known object observations used to calculate the PSF. Therefore, to enhance the quality of the PSF maps, we calculate them using multiple observations of a known source. Thus, in this study we combine 24 different Tau A observations to evaluate the PSF at zenith (with respect to the array) for frequencies 30--45~MHz. The obtained PSFs are then averaged for each individual frequency, yielding combined PSFs of the instrument (Figure~\ref{f-psf}). 

The effective average PSF of LOFAR consists of the bright mainlobe and a series of sidelobes, forming an approximately hexagonal pattern. The sidelobes become more prominent with increasing frequency and as expected, their distance from the mainlobe centroid varies with frequency approximately as $1/f$. The sidelobe pattern appears to be asymmetric: sidelobes located at azimuths 270$\deg$--360$\deg$ are bright, while sidelobes located at azimuths 90$\deg$--180$\deg$ are practically impossible to detect. 

Most importantly, the effective PSF appears to be significantly bigger than the nominal PSF for this mode of observations. Thus, at 30~MHz the area within the half-maximum contour of the nominal PSF corresponding to zenith is 110~arcmin$^2$, while for the effective PSF derived using Tau A observations it is about 170~arcmin$^2$. 

This difference is also demonstrated by Figure~\ref{f-nominal}, which shows Tau A observed by LOFAR at 30MHz at 07:00~UT on 13 July 2017 compared with the nominal PSF of the instrument calculated for the same frequency, time, and $\alpha$ and $\delta$ coordinates. Since Tau A is significantly smaller than LOFAR PSF at this range of frequencies, one can expect the two maps to be very similar. Clearly, they are not: the mainlobe area of the empirically-derived PSF is about 1.6 times bigger than the mainlobe area of the nominal PSF. Furthermore, unlike the empirically-derived PSF, the nominal PSF shows much fainter, symmetric sidelobes.

\subsection{Ionospheric shift}
\label{s-ionos}

Metric radio emission is strongly affected by refraction in the Earth's ionosphere, resulting in a shift of apparent positions of radio sources. These shifts can be considered as a sum of constant and variable components, with the latter varying on a range of timescales from minutes to months.

Comparison of the apparent location of Tau A with its known coordinates makes it possible to evaluate the shift for different frequencies and different elevations.  Figure~\ref{f-ionos}(a) demonstrates the effect of average ionospheric shift. As expected, it quickly increases with zenith distance: at 30~MHz the ionospheric shift is smaller than 0.03$\deg$ (100~arcsec) at about 60$\deg$ over horizon, increasing to almost 0.4$\deg$ at about 15$\deg$ over horizon.

The average shift should scale with the radio-frequency as $\sim f^{-2}$, and hence, can be written as $\Delta z_{ir} (f,z) = f^{-2} \mathcal{R}(z)$. Fitting an analytical function to the position deviations normalised  by $\sim f^{-2}$ (shown as a black line in Figure~\ref{f-ionos}a) provides the best fit for the function $\mathcal{R}(z)$, yielding an empirical formula for the shift: 
\begin{equation}
\Delta z_{ir} (z) = \frac {5.59}{f^2}\left[13.01 \exp\left(\frac{z-44.417}{17.313}\right)-1\right],
\label{eq-refrac}
\end{equation}
where $z$ and $\Delta z_{ir}$ are in degrees and $f$ is in MHz. This formula is used to correct the observed intensity maps for the average ionospheric refraction shift.

Figure~\ref{f-ionos}(a) represents the constant component of the ionospheric refraction shift for June-September season.
However, the variable component of refraction is important too. Figure~\ref{f-ionos} (b-c) demonstrates the effect of refraction on the apparent position for Tau~A observed about 60$\deg$ over horizon at the frequency of 30~MHz. The variations with an amplitude of about 100~arcsec and a timescale of about 50--100~s most likely represent fast variations in the Earth's ionosphere. Variations on longer time-scales can be significant, with 5--20~min variations caused by ionospheric gravity waves being most significant, with amplitudes of arcminutes \citep{meja97}. 

In order to correct solar source positions for the effect of the variable refraction, one needs nearly simultaneous observations of several known bright objects with the positions in the sky relatively close to the Sun. Since finding such objects in the sky is nearly impossible, we do not correct for the effect of these slower ionospheric variations and therefore, they contribute to the position uncertainty. This uncertainty is represented by deviations of the observed positions from the average value (black line in the Figure~\ref{f-ionos}a). For the considered frequency range this uncertainty is about 100--150~arcsec \citep{gore19}.

Figure~\ref{f-ionos} (b-c) also reveals fast oscillations with the amplitude of about 30~arcsec on the timescale of few seconds. These variations represent the error of centroid position measurements determined primarily by the characteristics of the TAB array used for observations. Since they occur on periods shorter than the integration time, they enhance the apparent area of the instrument's PSF and solar sources and therefore, are accounted for during the deconvolution procedure.

\begin{figure*}[ht!]
\centering{\includegraphics[width=0.6\textwidth]{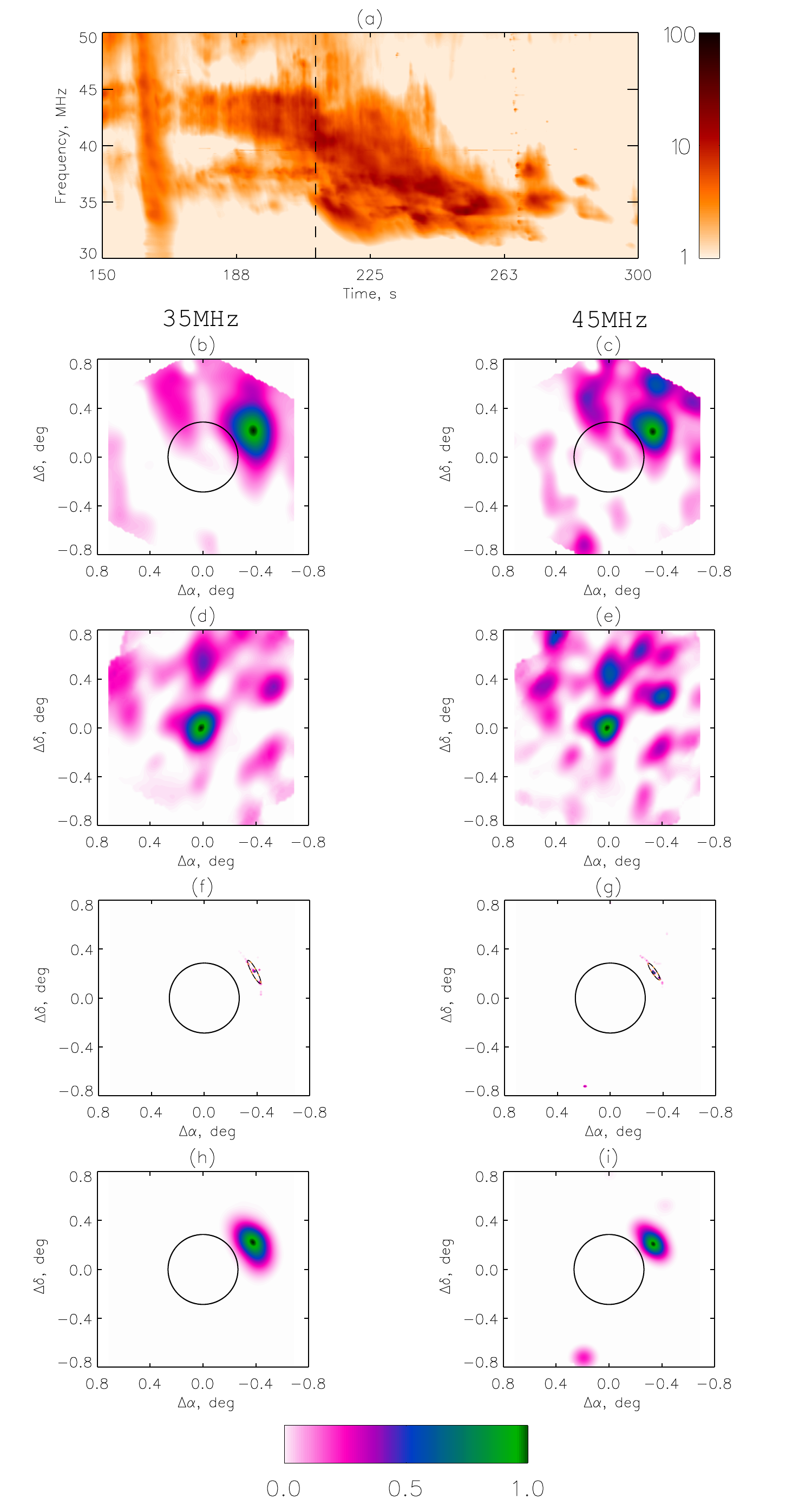}}
\caption{Solar radio map deconvolution procedure. Panel (a) shows the dynamic spectrum, the logarithm of intensity versus time and frequency, for the considered event (a ``transitioning'' Type II burst, reported by \cite{chre20}), with X-axis showing time in seconds after 10:59:38~UT on 15 July 2017. Panels (b) and (c) show dirty maps observed at 11:03:07~UT (shown as black dashed line in panel (a)). Panels (d) and (e) show corresponding LOFAR PSFs derived using the combined PSF for the location of the Sun at the moment of observations. Panels (f) and (g) show corresponding clean component maps (colour scales, Sect.~\ref{s-solar}) and their 2D Gaussian fits (black and orange dashed lines show their half-maximum contours, Sect.~\ref{s-fitting}). Panels (h) and (i) show corresponding cleaned maps (as described in Section~\ref{s-solar}). Panels (b), (d), (f), and (h) are for 35MHz; panels (c), (e), (g), and (i) are for 45MHz.}
\label{f-clean}
\end{figure*}

\section{Sizes and shapes of solar radio sources}
\label{s-results}

\subsection{Deconvolution of solar radio images}
\label{s-solar}

\begin{figure}[ht!]
\centering{\includegraphics[width=0.38\textwidth]{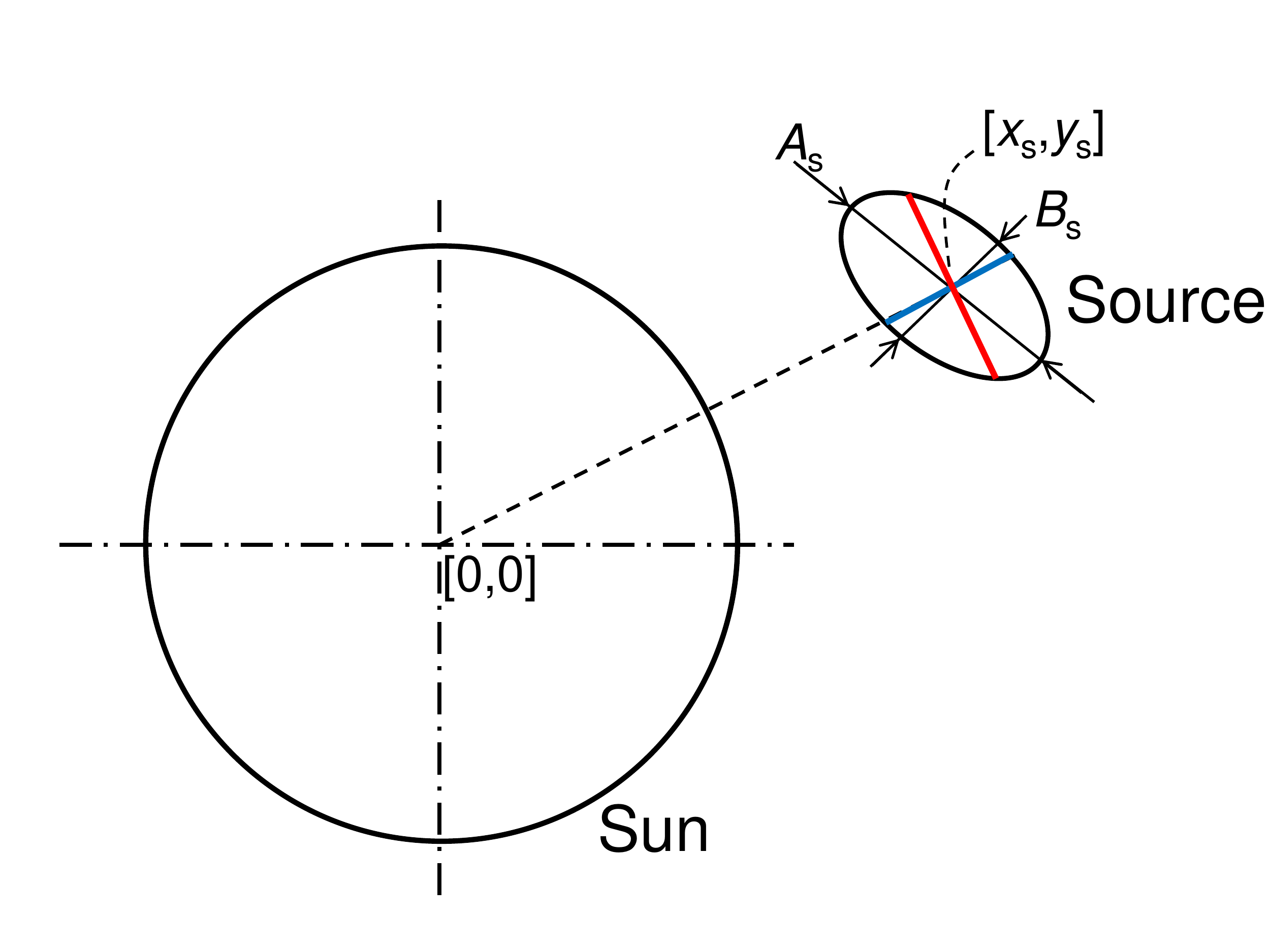}}
\caption{The sketch showing how the radial (blue line) and tangential (red line) sizes are defined, based on the location, and major and minor axes of the ellipse, corresponding to the half-maximum (HM) contour of the 2D Gaussian representing the intrinsic source.}
\label{f-lines}
\end{figure}

\begin{figure}[ht!]
\centering{\includegraphics[width=0.48\textwidth]{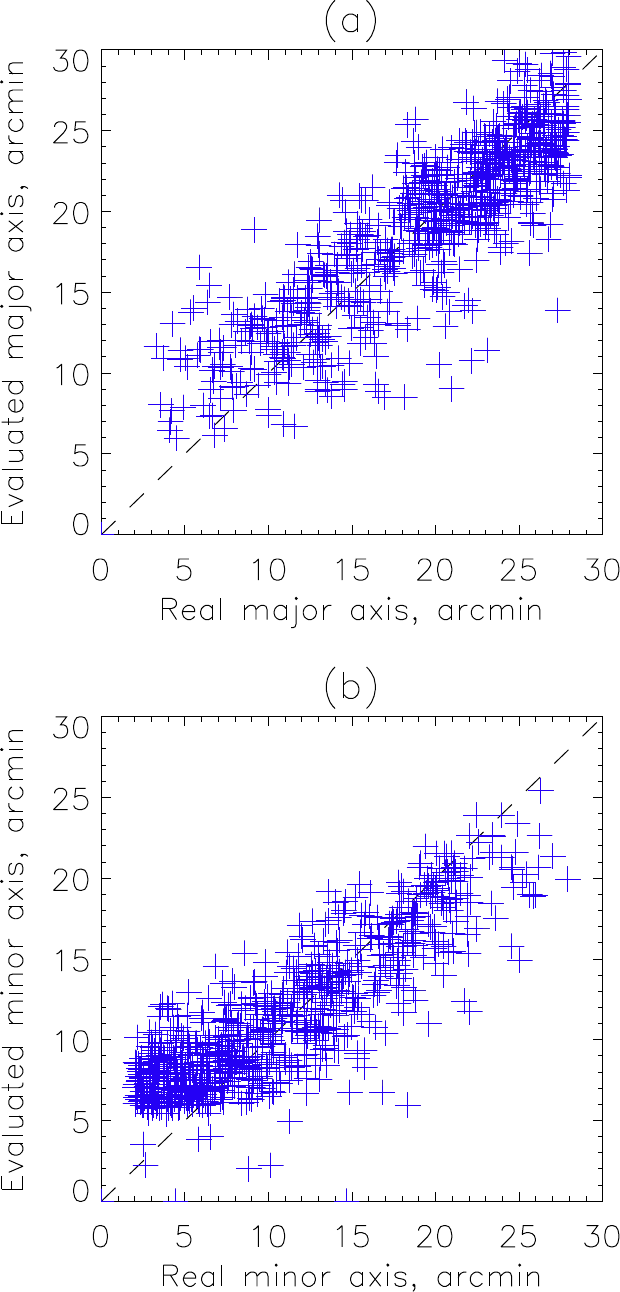}}
\caption{Comparison of the evaluated synthetic source sizes, major and minor axes of the HM ellipses, with their actual values. The sizes were evaluated using the same set of procedures as used for solar data processing.}
\label{f-error}
\end{figure}

\begin{figure}[ht!]
\centering{\includegraphics[width=0.48\textwidth]{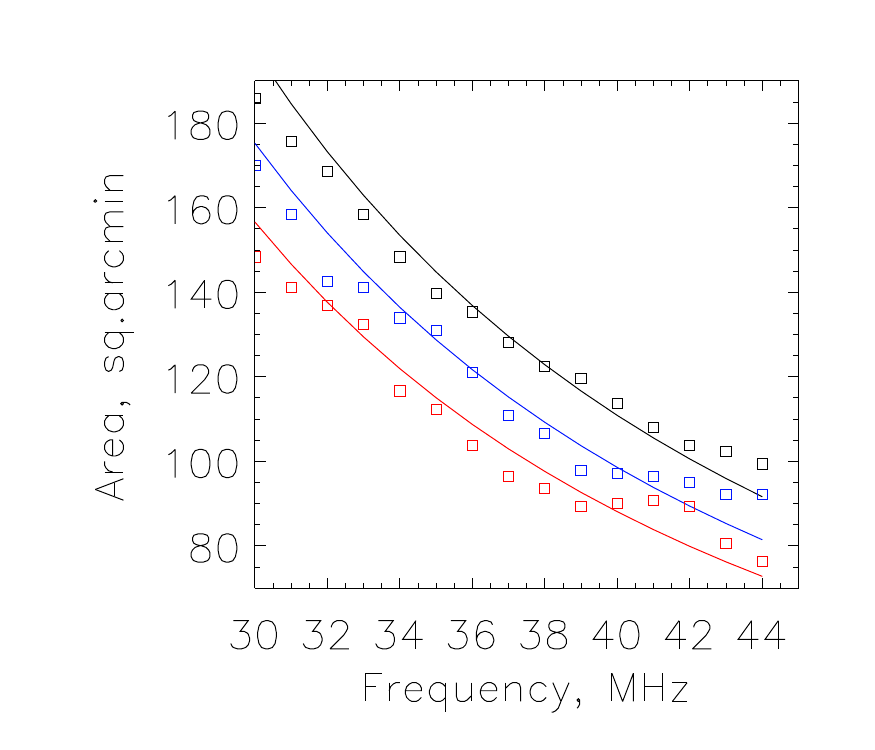}}
\caption{Area of the effective LOFAR PSF at zenith evaluated using different estimates for the constant background intensity. Black squares are for the PSF calculated with $N=2$, blue and red squares are for $N=109$ and $N=200$, respectively. Lines with corresponding colours show $1/f^2$ data fits.}
\label{f-area}
\end{figure}

The image deconvolution procedure, or cleaning, for each individual solar observation at each frequency consists of two parts: calculation of the PSF and then actual deconvolution. The PSF is calculated using the combined PSF discussed in Section~\ref{s-psf}. For each frequency the PSF is translated from zenith to the sky coordinates $(A,z)$, corresponding to the location of the Sun at the considered time using Equations~(\ref{eq-psf1}--\ref{eq-psf4}). The coordinates of the map are corrected for the effect of ionospheric refraction: its equatorial coordinates are shifted so that its zenith distance $z$ increases by the value calculated using Equation~(\ref{eq-refrac}). 

Deconvolution of the observed solar intensity maps (or {\it cleaning}) is done using the CLEAN algorithm \citep{hogb74,hure02}. This algorithm is based on the assumption that the real intensity map (or {\it real image}) can be represented by a linear combination of a variable background and a set of point-like sources. Convolved with the PSF, the real image yields the observed intensity map, or `dirty image'.  The CLEAN algorithm iteratively calculates the locations and intensities of the point-like sources, producing the {\it clean component map} and the residual variable background map. Since the map of clean components is noisy (which is inevitable since deconvolution is a classical ill-posed problem), the result of the clean component map is usually convolved with the {\it clean beam}, which usually represents the mainlobe of the corresponding PSF with sidelobes removed. The clean components map convolved with the clean beam and added to the map of residuals yields the {\it clean image}, and is done purely for representation purposes. Therefore, the difference between the dirty maps and clean maps produced by the CLEAN procedure is that the clean maps are free from artefacts caused by the sidelobes, while the apparent sizes of sources remain practically the same.   

Deconvolution for one of the considered events is shown in Figure~\ref{f-clean}. It starts with the dirty images with the constant background subtracted (Figure~\ref{f-clean} b-c). The PSF for the event (Figure~\ref{f-clean} d-e) is derived using the combined PSF, as described in Sect.~\ref{s-psf}. The CLEAN procedure yields the clean component maps  (Figure~\ref{f-clean} f-g). The clean images are obtained by convolving the clean components with the clean beam, which is derived by fitting a 2D Gaussian to the mainlobe of the PSF (Figure~\ref{f-clean} h-i).

Since our aim is to evaluate the intrinsic sizes of the sources, we are interested in the clean component maps, rather than the clean images. In order to evaluate the sizes and shapes, the clean component maps are fitted with 2D Gaussian distributions for each frequency. This is done using the regularisation procedure described below. 

\subsection{Fitting the clean component}
\label{s-fitting}

Since the map of clean components is very noisy, the fitting algorithms often fail to converge or produce obviously false results in most cases. In order to overcome this problem, we introduce regularisation into clean component fitting as follows. Firstly, we convolve the map of clean components with $\mathcal{F}_r$ -- a 2D Gaussian distribution with the diameter of HM contour $w_r$. The brightest source at the obtained map is then fitted with $\mathcal{F}_c$ -- a 2D Gaussian distribution, located at $[x_c,y_c]$ with major and minor axes of its HM ellipse $A_c$ and $B_c$, respectively, and tilt $\theta_c$, the angle between the major axis and X-axis measured clockwise. Since $\mathcal{F}_c$ represents convolution of $\mathcal{F}_r$ with the intensity distribution in the intrinsic source $\mathcal{F}_s$, and $\mathcal{F}_c$ and $\mathcal{F}_r$ are 2D Gaussians, $\mathcal{F}_s$ is also 2D Gaussian and its parameters can be evaluated using the well-known Gaussian convolution rule. Hence, $\mathcal{F}_s$ is a 2D Gaussian with major and minor HM axes
\begin{eqnarray}
A_s &=& \sqrt{A_c^2 - w_r^2}\\
\label{eq-size1}
B_s &=& \sqrt{B_c^2 - w_r^2},
\label{eq-size2}
\end{eqnarray}
respectively, with the same centroid location, $[x_s,y_s]=[x_c,y_c]$, and same tilt angle of its HM ellipse, $\theta_s=\theta_c$. 

Similar to most other studies, by the shape of a source we mean the ellipse  corresponding to 0.5 of the intensity of the fitted 2D Gaussian distribution. Using the obtained fit parameters, we calculate the average sizes of the sources as
\begin{equation}
L_a = \sqrt{A_s B_s}.
\end{equation}
The sizes of sources in radial direction (i.e. measured along the line connecting the source centroid and the centre of the solar disk, Figure~\ref{f-lines}) and tangential directions (i.e. perpendicular to the radial directions) are calculated as
\begin{eqnarray}
L_r &=& \frac{A_s B_s}{\sqrt{(A_s \sin(\theta_s-\lambda))^2+((B_s \cos(\theta_s-\lambda))^2}}\\
L_t &=& \frac{A_s B_s}{\sqrt{(A_s \cos(\theta_s-\lambda))^2+((B_s \sin(\theta_s-\lambda))^2}},
\label{eq-size3}
\end{eqnarray}
respectively, where $\lambda=\arctan(y_s/x_s)$.

\subsection{Uncertainty of solar source size measurements}
\label{s-error}

The uncertainties in the source sizes evaluated using the approach described in Section~\ref{s-fitting} have been estimated using synthetic data. A large number ($10^3$) of synthetic sources with 2D Gaussian distributions were generated with randomly chosen centroid locations, major and minor HM axes, and tilt angles (angles between the major axes and horizontal axis). The synthetic sources were convolved with the actual LOFAR PSF, and reduced on an irregular grid identical to the LOFAR beam grid, with random noise ($\sim 0.1$ of the source amplitude) and constant background ($\sim 10$--$20$ of the source amplitude) added, creating synthetic `dirty maps'. The obtained `dirty maps' were then analysed using the same procedure as described in Section~\ref{s-data} and Sections~\ref{s-solar}--\ref{s-fitting}. By comparing the derived parameters of the sources with their actual parameters (Figure~\ref{f-error}) it was found that the error in minor and major axes measurements is 3~arcmin. For an average source considered in this study, this error in size measurements would translate to an uncertainty in tilt angle measurements of about 15--20~degrees. Although the error in centroid position estimated using the synthetic sources is only 1~arcmin, in reality it is substantially larger, around 3--5~arcmin, due to the variable ionospheric refraction (see Section~\ref{s-ionos}, also \cite{gore19}).  

The constant background level estimation (Section~\ref{s-obs}) is another source of uncertainty. The assumption that at least 50\% of the field of view is free from `real' signal might be incorrect for large sources. Although this assumption should affect the synthetic source measurements done above, i.e. the error due to background uncertainty is included into the error estimated above, we evaluate separately the error due to the background level uncertainty. This is done by measuring Tau A areas assuming that only a small fraction of the field of view ($N=2$) is free from signal, and assuming that nearly the whole field of view is free from signal ($N=200$). As expected, the measured areas decrease with increasing $N$. However, this difference is moderate: for one example, the effective PSF area measured at 30~MHz with $N=2$ is about 18~arcmin$^2$ ($\sim$10\%) bigger than the area measured with $N=109$, while the area measured with $N=200$ is approximately 17~arcmin$^2$ smaller than the area measured with $N=109$ (Figure~\ref{f-area}). The resulting error in the measurements of solar source sizes caused by the uncertainty in the background intensity value is about 1.8~arcmin i.e. significantly lower than the overall error ($\sim$2.5-3~arcmin). This is because the background intensity is evaluated using the same approach and the same value of $N$ both for Tau A and solar intensity maps, and the systematic errors in solar and Tau A measurements partially cancel each other.     

\subsection{Solar source sizes at different frequencies}
\label{s-sizes}

\begin{figure*}[ht!]
\centering{\includegraphics[width=0.98\textwidth]{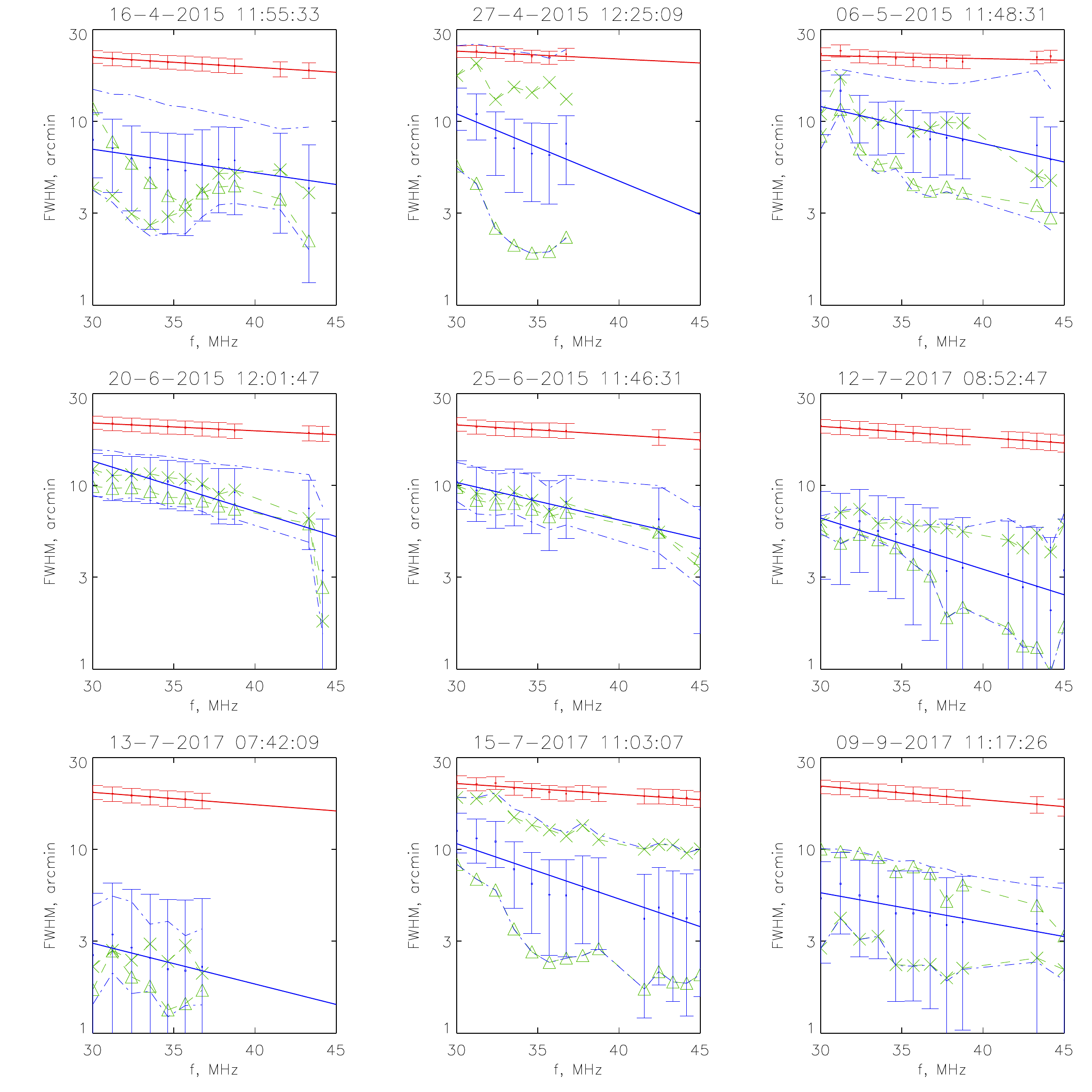}}
\caption{Sizes of solar radio sources observed at different frequencies by LOFAR in nine randomly selected separate events (dates and UT times are shown above the panels). Red dots with error bars show the average sizes of the brightest sources on dirty maps, while solid red lines show power-law fits ($a f^{-\gamma}$). Similarly, blue dots with error bars show the average sizes of intrinsic sources, along with the power-law fits (solid blue lines). Blue dot-dashed lines show the sizes of major and minor axes of the half-maximum ellipses corresponding to intrinsic sources. Green dashed lines with crosses and triangles show sizes of intrinsic sources in tangential and radial directions, respectively.}
\label{f-sizes}
\end{figure*}

The resulting sizes of the observed solar sources are shown in Figure~\ref{f-sizes}. It can be noticed that at some frequencies, particularly above 40~MHz, the data is missing because the 2D Gaussian fitting procedure failed to converge. This may have happened for a number of reasons, including low signal, high noise level, or radio-interference. However, where the data exists, it shows very similar patterns in most events.

The evaluated source sizes can be very different. Thus, at 30~MHz, the average sizes in the considered events range from 3 to about 15~arcmin, although the majority of them (seven out of nine) are in a much smaller range between 8--12~arcmin. One of the events (type III burst on 13/07/2017) has very small average size of about 3~arcmin (i.e. comparable to the error of size measurements in this approach) at 30~MHz. At the same time, in one case (type IV event on 20/06/2015) the average source size at this frequency is about 15~arcmin. In all considered events, the  average sizes clearly decrease with increasing frequency, and can be approximated by a power-law $\sim f^{-\gamma}$. For the majority of events, $\gamma$ is between 1--2.

The evaluated shapes of solar radio sources show significant ellipticity, with the sizes of major axes being about a factor of 3 larger than the corresponding minor axis sizes. In seven out of nine considered events, the radial sizes are significantly smaller than the tangential sizes. At 30~MHz, the tangential sizes (which often nearly coincide with the major axes of the fitted Gaussians) are typically 10--20~arcmin, while the sizes measured in radial direction are typically 5--10~arcmin.  

\section{Discussion and conclusions}

In this study, we have achieved two major goals. 

Firstly, we have developed and tested a novel, simple method for deriving the PSF of an instrument based on known nearly-point-like source observations at an arbitrary location. The derived method can be used for deconvolving intensity maps of the Sun or other bright radio objects, using a limited pool of calibration sources. This will enable an accurate analysis and interpretation of imaging observations obtained using instruments operating in TAB mode. We showed that the presented method works well, even at higher frequencies where LOFAR's TAB mode PSF has a rather complex structure. However, our tests also indicated that the known sources used for evaluating the instrument's PSF need to be observed with a field of view about twice as large as the field of view of the considered solar observations, in order to minimise the artefacts present near the field of view edges in the cleaned images. 

The difference between the nominal and effective LOFAR PSFs is significant. In the 30--45~MHz range, the area of the effective PSF is about 1.6 times bigger than the nominal PSF area. Hence, using the nominal instead of the effective PSF may result in a significant overestimation of the intrinsic emission source sizes. Therefore, taking into account the unique role LOFAR is expected to play in the next few years in the exploration of the upper solar corona, evaluation of the effective LOFAR PSF and the PSF translation method are very important results in their own right. Furthermore, the developed method can be used for the analysis of imaging observations with future instruments, such as the Low-Frequency Aperture Array of the Square Kilometer Array (SKA).

Secondly, for the first time, we have evaluated the intrinsic sizes and shapes of solar metric emission sources and their variation with frequency in the range 30--45~MHz. At 30~MHz, in seven out of nine events the average source sizes are between 8--12~arcmin, which is around 2--4 times smaller than some previous observations, including observations with LOFAR. In seven out of nine considered cases, the sources appear to be significantly smaller in the solar radial direction.

It is important to note that the considered sources of radio-emission originate in radio bursts of different types. Although we believe that the observed emission is produced by the plasma mechanism, because of its narrow bandwidth ($\Delta f / f \sim 0.1$), the physical sizes of the regions producing the observed emission, as well as the mechanisms of electron acceleration in these events are expected to be very different. Thus, type II and IV bursts are expected to be produced by large coronal structures, while physical volumes producing type III emission are expected to be relatively small, comparable to the cross-section of a magnetic flux-tube in the upper corona \citep{kone17,muse21}. However, although out of nine considered events the smallest size corresponds to a type III burst (13 July 2017 event, Figure~\ref{f-sizes}), while the largest (in terms of its average size) source corresponds to the type IV burst (20 June 2015 event), there is no systematic difference between type II and IV, and type III events. 

Hence, at least for the considered Type III events, one can say that the source sizes in solar radio bursts -- even when corrected for the effective PSF -- are much larger than the expected intrinsic sizes of the emission sources. At the same time, the source sizes, their variation with frequency, and most importantly, ellipticities of the sources are consistent with the models of anisotropic radio-wave scattering in the corona and inner heliosphere \citep{kone19,kuze20,chee20}. Therefore, our findings support the concept that the sizes of low-frequency solar radio-sources are determined primarily by radio-wave scattering in the upper corona, rather the physical sizes of the emitting regions.

\section*{Acknowledgements}
MG and PKB were supported by the Science and Technology Facilities Council (STFC, UK), grant ST/T00035X/1. DLC and EPK are thankful to DSTL for the funding through the UK-France PhD Scheme (contract DSTLX-1000106007). EPK was supported by STFC consolidated grant ST/P000533/1. NC thanks CNES for its financial support. The authors acknowledge the support by the international team grant (http://www.issibern.ch/teams/lofar/) from ISSI Bern, Switzerland. 
This paper is based (in part) on data obtained from facilities of the International LOFAR Telescope (ILT) under project code LC3\_012, LC4\_016, and LC8\_027. LOFAR \cite{vane13} is the Low Frequency Array designed and constructed by ASTRON. It has observing, data processing, and data storage facilities in several countries, which are owned by various parties (each with their own funding sources), and that are collectively operated by the ILT foundation under a joint scientific policy. The ILT resources have benefited from the following recent major funding sources: CNRS-INSU, Observatoire de Paris and Universit\'e d'Orl\'eans, France; BMBF, MIWF-NRW, MPG, Germany; Science Foundation Ireland (SFI), Department of Business, Enterprise and Innovation (DBEI), Ireland; NWO, The Netherlands; STFC, UK; Ministry of Science and Higher Education, Poland.

\bibliographystyle{aasjournal}
\bibliography{na21}

\end{document}